\documentclass[11pt]{article}

\usepackage{amsmath}
\usepackage{amssymb}

\usepackage{lineno}

\usepackage{color, colortbl, array}
\definecolor{Gray}{gray}{0.9}

\usepackage{umoline}
\usepackage{dcolumn}
\usepackage{color}

\usepackage{esvect}

\usepackage{paralist}
\usepackage[subnum]{cases}

\usepackage{graphicx}
\DeclareGraphicsExtensions{.pdf}

\usepackage{epstopdf}

\usepackage{paralist} 

\usepackage{cite}

\usepackage{color} 

\usepackage[subnum]{cases}

\usepackage[labelfont=bf,labelsep=period,justification=raggedright]{caption}

\makeatletter
\renewcommand{\@biblabel}[1]{\quad#1.}
\makeatother

\date{}

\begin{document}


\begin{center}

{\Large

\textbf{On the scaling patterns of infectious disease incidence in cities }
}
\\
\vspace{0.2in}
 Oscar Patterson-Lomba$^{\ast 1}$, Andres Gomez-Lievano$^{2}$
\\
\vspace{0.2in}
$^{1}$Analysis Group Inc., Boston, MA, USA\\
$^{2}$Center for International Development, Harvard Kennedy School, Harvard University, Cambridge, MA, USA\\

$\ast$E-mail: pattersono85@gmail.com
\end{center}

\begin{abstract}
Urban areas with larger and more connected populations offer an auspicious environment  for contagion processes such as the spread of pathogens. Empirical evidence reveals a systematic increase in the rates of certain sexually transmitted diseases (STDs) with larger urban population size. However, the main drivers of these systemic infection patterns are still not well understood, and rampant urbanization rates worldwide makes it critical to advance our understanding on this front. Using confirmed-cases data for three STDs in US metropolitan areas, we investigate the scaling patterns of  infectious disease incidence in urban areas. The most salient features of these patterns are that, on average, the incidence of infectious diseases that transmit with less ease-- either because of a lower inherent transmissibility or due to a less suitable environment for transmission--  scale more steeply with population size, are less predictable across time and more variable across cities of similar size. These features are explained, first, using a simple mathematical model of contagion, and then through the lens of a new theory of urban scaling. These theoretical frameworks help us reveal the links between the factors that determine the transmissibility of infectious diseases and the properties of their scaling patterns across cities. 

\end{abstract}

\tableofcontents

\section{Introduction}

Urban populations throughout the world are growing rapidly, with predictions indicating that by mid-century over 60\% of the population will live in urban areas \cite{urbanUN}. This explosion of urban dwellers makes it paramount to decipher how this urbanization transition can be sustainable and beneficial at a global scale. Crucial to this enterprise is to improve our understanding of how infectious diseases spread and evolve in cities, as well as the overall public-health implications of  rampant urbanization \cite{ aparicio2002markers, aparicio2009mathematical, zhang2008modelling, alirol2011urbanisation, galea2002urbanization, cohen2010urbanization, harpham2001urban, leon2008cities, galea2005cities, bamaiyi2013role, lonnroth2009drivers}.

By their nature, cities foster large populations living in high proximity, leading to high levels of human interaction and circulation. Cities are thus suitable landscapes for pathogens to disseminate locally and internationally  \cite{clark2010germs,morens2013emerging,  alirol2011urbanisation, bettencourt2011bigger}. Today, the rapid urban population growth and density in developing countries, combined with poor living conditions and a precarious public health infrastructure, offer a particularly favorable landscape for the spread of certain infectious diseases.

 Increased population density and urbanization may have played a critical role in the worldwide dissemination of HIV \cite{quinn1994population, cohen2010urbanization, bettencourt2007growth}, the worsening of epidemics of major respiratory viruses (e.g., Influenza, RSV) \cite{glezen2004changing}, as well as the incidence of tuberculosis \cite{alirol2011urbanisation, stephens1996healthy, lonnroth2009drivers, aparicio2002markers}. Moreover, the rates of STDs are, on average, higher in urban areas as compared to the national rates in the U.S \cite{patterson2015per}. Socio-economic factors such as education, healthcare accessibility, income and  social inequalities have been identified as relevant to the spread of STDs, but causal links are difficult to establish \cite{diclemente2005prevention, holtgrave2003social, farley2006sexually, crosby2003social, semaan2007social, patterson2015per}. 

Some of the efforts trying to provide a mechanistic understanding of the dynamics of infectious diseases in cities have focused on a particular disease in one city at a time \cite{alirol2011urbanisation, mao2010spatial, gesink2011sexually}. However, this approach can often lead to models that are context (city/region) specific and too mathematically complex to tease out generalizable insights. Here, we present some empirical observations and analytical arguments that hold promise for a generalizable understanding of the consequences of urbanization on the spread of diseases.

\section{Empirical findings}

\begin{figure*}[h!]
\begin{center}
 \includegraphics[width=0.7\textwidth, trim = 0in 0.3in 0in 0.5in,clip=true]{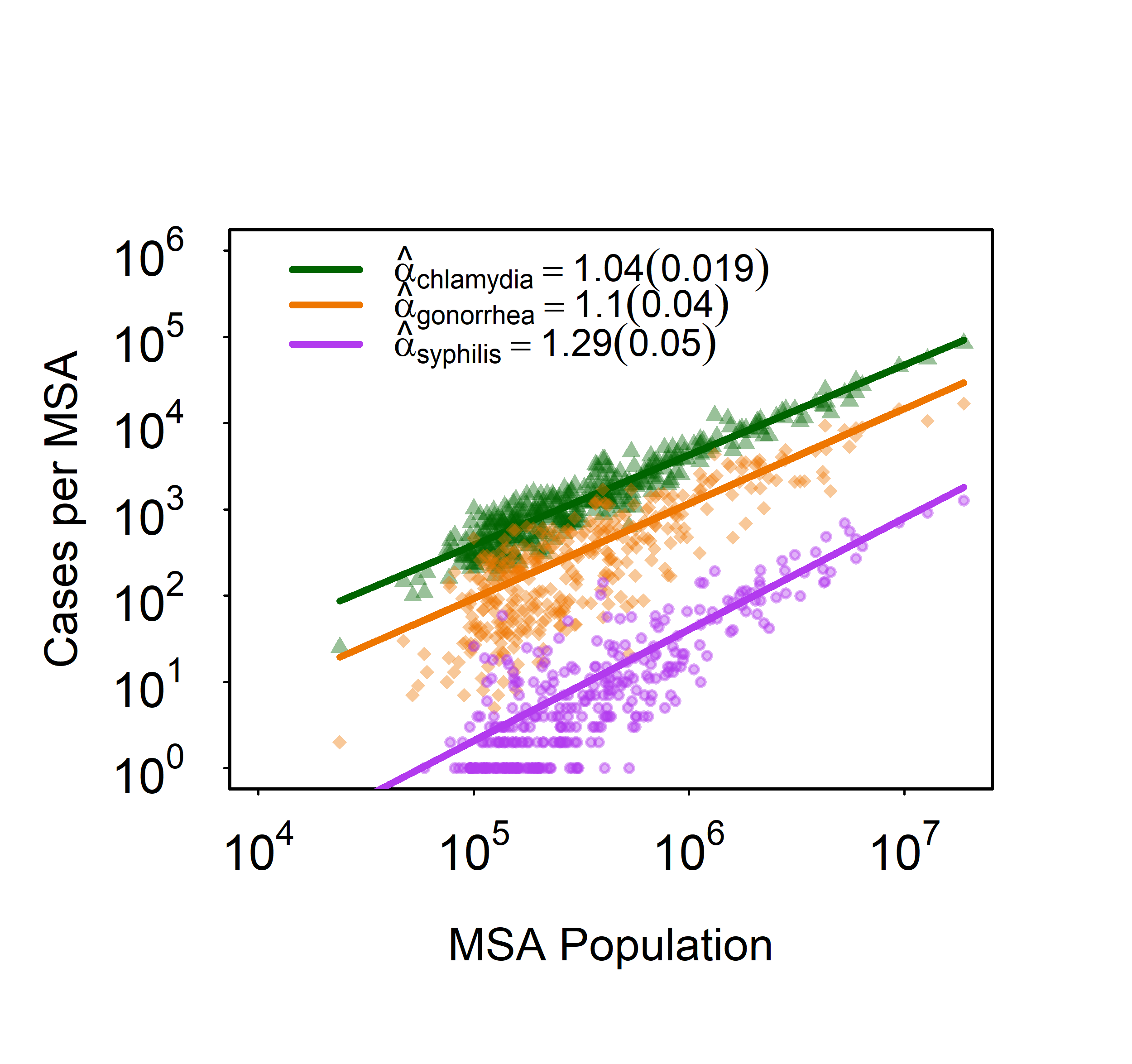}
\caption{\small Scaling of  chlamydia, gonorrhea and syphilis incidence with MSA population with Negative Binomial regression lines using model \eqref{logyn}. The $\hat{\alpha}$ estimates represent the slopes of the regressed lines. An estimate greater than 1 suggest a \emph{superscaling} pattern.  }
\label{NB_chlam}
\end{center}
\end{figure*}

Focusing on confirmed case-data for chlamydia, gonorrhea and syphilis from 2007 to 2011 \cite{cdcSTD12}, we previously performed a cross-sectional analysis of disease incidence in Metropolitan Statistical Areas (MSAs) of the U.S. and identified interesting statistical patterns at a \emph{systemic} level \cite{patterson2015per}. The results, depicted in  Figure \ref{NB_chlam},  suggested that the per-capita incidence of these STDs increases systematically with the population size of MSAs, even after controlling for important socio-economic covariates. In addition, we identified significant differences in the scaling patterns of these three STDs. Similar findings were recently reported regarding infectious diseases such as HIV, Influenza and Meningitis in Brazil \cite{rocha2015non}. However, the main drivers of these systemic infection patterns are still not well understood. 

Identifying these drivers is important to, for instance, help us understand to what extent the high incidence of an infectious disease in a city is due to a suitable environment (poor public health infrastructure), or the specific characteristics of the disease (asymptomaticity), or the behaviors of its residents (high frequency of personal interactions).

\subsection{Extended Statistical Model}

Do the scaling features of STDs depend  on what ``type" of MSAs we focus on? To address this question we investigate in more detail  how the associations of urban population size and the transmission dynamics of infectious diseases depend on properties of the disease \emph{and} on the socio-economic context of the populations in which they spread. This in turn will  shed light on the potential drivers of the scaling patterns of STDs.
 
To that end, we introduce an extension of the classical statistical framework in \cite{bettencourt2007growth, patterson2015per}.  Briefly, this classical framework posits that $Y$, an urban metric, and $N$, the population size of an urban area, satisfy the scaling relationship: 
\begin{equation}
\mathbb{E}[Y|N_i]=Y_0N_i^{\alpha}. \label{yn}
\end{equation}

\noindent  The subscript indexes a city or urban population, $Y_0$ is a baseline value for $Y$, and the scaling exponent $\alpha$ measures the average relative change in $Y$ with respect to $N$. The scaling relationship  in Eq. (\ref{yn}) can be transformed to:
\begin{equation}
\log(\mathbb{E}[Y|N_i])=\log(Y_0)+\alpha \log(N_i), \label{logyn}
\end {equation}

\noindent from which empirical regularities can then be investigated by estimating the scaling exponent $\alpha$ and the baseline prevalence, $Y_0$. Using \eqref{yn} it is easy to show that when $\alpha>1$ (\emph{superscaling}), the expected incidence rate given by $Y/N=Y_0 N^{\alpha-1}$ is an increasing function of $N$. In other words, when the scaling relation is \emph{superlinear}, infection \emph{rates} increase with population size.

 To better understand how socioeconomic properties can impact the scaling profiles for each disease, and thus the nature of the scaling patterns and its drivers, we explore the potential effects of certain economic and demographic factors in the scaling patterns (i.e., as effect modifiers of $\alpha$ and $Y_0$).

Hence, we modify the original scaling model in \eqref{yn} and test whether or not the scaling exponent $\alpha$, as well as the intercept $Y_0$, depend on the values of the socio-economic covariates.  The modified model is given by: 
\begin{eqnarray}
\mathbb{E}[Y|N_i]=Y_{0j} e^{\xi_{j} X_{ij}}N_i^{\alpha_{0j}+\alpha_{1j} X_{ij}} \label{interact}
\end{eqnarray}

where, as before, $Y$ is the incidence of cases in a given MSA with population size $N_i$, and $X_{ij}$ is the value of the covariate $j$ in MSA $i$, with corresponding coefficients $\alpha_{0j}$ and $\alpha_{1j}$ (for the exponent), and $\xi_{j}$ (for the intercept). The six covariates we explore here are:  \emph{\% African American}, \emph{Gini index}, \emph{\% Poor}, \emph{education index},  \emph{income per-capita}, and  \emph{\% insured}. The choice of these particular socio-economic covariates reflects what previous studies have found to be key factors in the spread of STDs \cite{holtgrave2003social, farley2006sexually, semaan2007social, marmot2005social,  patterson2015per}.

The modifications in the scaling equation \eqref{interact} affect both the intercept, $Y_0^{all}=Y_{0j} e^{\xi_{j} X_{ij}}$, and the scaling exponent $\alpha^{all}=\alpha_{0j}+\alpha_{1j} X_{ij}$. To see this more clearly, taking logarithm in both sides of  \eqref{interact} yields

\begin{eqnarray}
\log(\mathbb{E}[Y|N_i])=\underbrace{\left(\log(Y_{0j}) +\xi_{j} X_{ij}\right)}_{Y_0^{all}} +\underbrace{(\alpha_{0j}+\alpha_{1j} X_{ij})}_{\alpha^{all}}\log(N_i) +\epsilon_i. \label{interact2}
\end{eqnarray}

Given that we are dealing with an overdispersed count variable, we regress model  \eqref{logyn} and \eqref{interact2} using Negative Binomial regression \cite{ismail2007handling} (see \cite{patterson2015per} for details).

Using model \eqref{logyn} we obtain $\alpha>1$ for all three STDs, indicating that the rates of STDs appear to increase in a systematic way with population size (superlinear scaling), as shown in Figure \ref{NB_chlam} and in Table \ref{table:NB_chlam}.

\begin{table}[h!] \centering 

\footnotesize 
 \begin{tabular}{| c|  c | c | c | }

\hline
Covariates &\textbf{Chlamydia} &\textbf{Gonorrhea} &\textbf{Syphilis} \\ 
\hline 
\hline
\rowcolor{Gray}
   $\widehat{\alpha}$ & 1.04 & 1.10 & 1.29   \\ 
 ($SE$)  & (0.019) & (0.040) & (0.051) \\ 
 \rowcolor{Gray}
 $\log(\widehat{Y_0})$ & -1.94 & -2.84 & -5.46 \\ 
($SE$)   & (0.105) & (0.219) & (0.280) \\ 
\hline 
\hline
Observations & 364 & 364 & 364 \\
\hline
\end{tabular} 
  \caption{\emph{Negative Binomial regression results in the four infectious diseases using model \eqref{logyn}.}} 
  \label{table:NB_chlam} 
\end{table}

Before delving into the results from the extended model, the six covariates explored here can be braodly categorized, based on their correlation with STD per-capita rates (see Table \ref{table:covariate_correlation}), as \emph{positively} correlated (\emph{\% African American}, \emph{Gini index} and \emph{\% Poor}) and \emph{negatively} correlated (\emph{Education index},  \emph{Income per-capita}, and  \emph{\% Insured}). For example, a city with a larger proportion of poor residents is expected, on average, to suffer from larger prevalence of certain STDs, whereas a city with a higher proportion of insured residents is, on average, better equiped to contain the spread of STDs. Note that these observations are based on correlations rather than on \emph{causal} relations.

\begin{table}[h!] \centering 

\footnotesize 
 \begin{tabular}{| c|  c | c | c | }

\hline
\textbf{Covariates} &\textbf{Chlamydia} &\textbf{Gonorrhea} &\textbf{Syphilis} \\ 
\hline 
\hline
\rowcolor{Gray}
   \% African American & 0.74 & 0.85 & 0.58   \\ 
 Gini index  & 0.25 & 0.28 & 0.31 \\ 
 \rowcolor{Gray}
 \% Poor & 0.37 & 0.31 & 0.17 \\ 
Education index   & -0.16 & -0.12 & -0.06 \\ 
 \rowcolor{Gray}
 Income per-capita & -0.20 & -0.17 & 0.01 \\ 
\% Insured   & -0.15 & -0.10 & -0.18 \\ 
\hline 

\end{tabular} 
  \caption{\emph{Pearson correlation coefficients corresponding to of each covariate with per-capita incidence (incidence/population) across cities. The first three covariates are positively correlated with per-capita incidence rates, whereas the last three are negatively correlated with per-capita incidence rates (except for the case of income and syphilis where the correlation is positive but very low).}} 
  \label{table:covariate_correlation} 
\end{table}

Table \ref{table:NB_chl2} shows the results of the extended model in \eqref{interact2} that accounts for interaction terms for each of the covariates separately (i.e., one covariate at a time). Of particular interest, are the estimates for $\alpha_1$ and $\xi$, which represent the effect that each covariate has, respectively, on the baseline prevalence and scaling exponent of the scaling relationships of each STD. 
Table \ref{table:NB_chl2} indicates that, overall, the \emph{positively} correlated covariates are found to reduce the scaling exponent (i.e., negative $\widehat{\alpha_1}$) while increasing the baseline prevalence (i.e., positive $\widehat{\xi}$), whereas   the \emph{negatively} correlated covariates are found to increase the scaling exponent and decrease the baseline prevalence. For the covariate \% Poor, for example, this pattern can be interpreted as: everything else equal, 1) the relative effect of population size on the number of STD cases is lower in MSAs with a high percentage of poor residents, and 2) MSAs with a high percentage of poor residents have larger baseline rates of STDs. That larger levels of the positively correlated covariates lead to higher levels of baseline prevalence is not surprising; however, that these larger levels are also associated with lower scaling exponents is not necessarily an intuitive result, nor is it one that can be fully or mechanistically understood using this statistical modeling approach.

 \begin{table}[h!] \centering 

\footnotesize 
\vspace{0.1in}
   
     \begin{tabular}{| c|  c |  c  | c | c | c | c | c |}
    \hline 
 \emph{\textbf{Disease}}    &   \emph{Covariates}     &$\widehat{\alpha_1} \ (p)$  &  $\widehat{\xi} \ (p)$    &  AIC diff.    \\ \hline
     \hline
 \rowcolor{Gray}   
\textbf{Chlamydia}  &    {\% African American}& -0.004 ( 0.003 )& 0.0334 ($<0.001$) & 231.13\\

& Gini & -1.647 ( 0.053 )& 10.758 ( 0.02 ) & 22.8  \\ 
   \rowcolor{Gray}

&\% Poor  & -0.005 ( 0.377 )     & 0.045 ( 0.138 )       & 68.3    \\

  & Education & -0.0143 ( 0.549 )& 0.034 ( 0.789 ) & 14.1 \\ 
   
 \rowcolor{Gray}
& Income & 3.174e-06 ( 0.408 )& -3.134e-05 ( 0.148 ) & 32.1 
\\ 
 
& \% Insured & 0.006 ( 0.085 )& -0.039 ( 0.044 ) & 9.4 
 \\ 
 \hline
 \hline
 \rowcolor{Gray}
\textbf{Gonorrhea} &    {\% African American} & -0.011 ($<0.001$)& 0.088 ($<0.001$)& 287.1 \\

& Gini & -2.759 ( 0.123 )& 19.062 ( 0.05 )& 24.7 \\ 
 \rowcolor{Gray}

&\% Poor & -0.019 ( 0.12 )& 0.141 ( 0.032 )&  57.2 \\

 & Education  & 0.025 ( 0.623 )& -0.216 ( 0.426 )& 9.6 \\ 
   
 \rowcolor{Gray}
& Income & 1.110e-05 ( 0.169 )& -9.311e-05 ( 0.041 )& 33.1 
\\ 
 
& \% Insured & 0.010 ( 0.18 )& -0.065 ( 0.119 ) & 2.6 
\\ 

\hline
\hline

\rowcolor{Gray}
\textbf{Syphilis} &   {\% African American} & -0.018 ($<0.001$)& 0.126 ($<0.001$)& 258.3 
 \\

& {Gini} & -5.551 ( 0.012 )& 40.133 ( 0.001 )& 82.7 
\\ 

 \rowcolor{Gray} 
&{ \% Poor} & -0.052 ( 0.001 )& 0.340 ($<0.001$)& 78.5 \\

&  Education & 0.071 ( 0.278 )& -0.516 ( 0.148 )& 18.4 
\\ 
   
  \rowcolor{Gray}
& {Income} & 2.642e-05 ( 0.012 )& -0.0002 ( 0.004 )& 19 
\\ 

& {\% Insured} & 0.023 ( 0.01 )& -0.164 ( 0.001 )& 41 
 \\ 
 
\hline 

    \end{tabular}
  \caption{\emph {Negative Binomial regression results for all STDs using model \eqref{interact2}. In parenthesis are the p-values testing the null that $\alpha_i=0$ and $\xi=0$. The AIC diff measures the difference in AIC between a model that only includes the population size as a predictor and a model also including the respective socio-economic covariate; a large positive difference indicates that the model with the added covariate is better at predicting  the response variable but yet parsimonious.} }
  \label{table:NB_chl2} 
\end{table}

The empirical findings paint a picture in which: i) different STDs feature significantly different scaling patterns (Figure \ref{NB_chlam} and Table \ref{table:NB_chlam}), and ii) the scaling features of STDs depend, on a systematic way, on the socioeconomic properties of the cities they spread in (Table \ref{table:NB_chl2}). In what follows, we provide interpretations for these observations using a mathematical epidemiological mode and  a novel theory of scaling.

\section{Explaining the scaling patterns}

Overall, population size has an enabling effect in the spread of STDs (at least the ones studied here). However, the scaling patterns for each STD differ in three important aspects. First, the scaling exponents are different \cite{patterson2015per}. Second, the log of the intercepts, $\log(Y_0)$, are also significantly different. And third, when comparing chlamydia, gonorrhea and syphilis the variance of the incidence given population size also differs significantly, with chlamydia having the least variance (as measured by the sums of residuals squared) and syphilis  having the highest variance. 

 Interestingly, diseases differ in these three properties in a systematic way. STDs with higher scaling exponent also have  lower intercept and lower variance given population size, and vice versa (see Figure \ref{NB_chlam}). Furthermore, the results  obtained using model \eqref{interact2} suggest that the scaling patterns also vary significantly when focusing on different types of MSAs (i.e., different socio-economic environments).

This methodology is, however,  unable to infer any causal relations between population size and ID incidence. Without a mechanistic understanding of these processes we are not capable of explaining important features of these patterns. For example, why do the scaling patterns differ between diseases? Or, why do the scaling exponents for the same disease can have contrasting results  in  different countries (as shown in \cite{rocha2015non}) or in different types of MSAs (as shown above)? These differences  suggest that intrinsic properties of the diseases and/or the  socio-economic landscape they spread in, may be driving disease spread in unique ways.

The superlinear scaling of the connectivity of the human contact networks with city size has been suggested as a mechanistic explanation for the superlinear patterns of infections across cities \cite{bettencourt2011bigger, rocha2015non}.  Indeed, recent empirical evidence from large mobile phone \cite{schlapfer2012scaling} and twitter  \cite{tizzoni2015scaling} datasets show that social connectivity scales superlineraly with population size. However, it still remains unclear how variations in the population size across cities, in combination with the non-linear effects of large social networks, can impact the scaling properties of epidemic outcomes.

To better make sense of the differences in  the scaling patterns of STDs noted above, we focus on the intrinsic transmission capacity of these STDs, and note that  symptomatology and inherent infectivity are key factors in the overall transmissibility of these pathogens. Chlamydia, having a  high infection risk per sexual act and being relatively asymptomatic (i.e., difficult to detect and prevent), features an advantageous combination of traits that makes it the most most transmissible of the four STDs studied herein. Syphilis, on the opposite end,  has low infectiousness and is the most symptomatic of the four STDs. As a result, chlamydia is accountable for a larger number of infections  as compared to less transmissible ones, such as syphilis and HIV. Presumably, these two propagate largely through high-risk (core) sexual networks \cite{hethcote1984gonorrhea} where individuals engage in riskier sexual practices. 

Meanwhile, the associations of incidence with population size (quantified by $\alpha$) decrease as STDs increase their transmissibility and prevalence. Therefore, the scaling analysis indicates that population size is more associated with the spreading capacity of STDs that transmit less readily.  According to this view, population size enables STD transmission, and more so for STDs that transmit with less ease. This means that the probability of contagion of low transmissibility diseases, like syphilis, are heavily dependent on the characteristics of the environment through which they can spread. This in turn implies that the probability syphilis has for spreading should be very sensitive to the connectivity patterns in a city, which is in turn a function of population size. Not surprisingly, STDs that transmit more easily are more prevalent and therefore are expected to have a higher intercept, thus the negative correlation  between scaling exponents and  intercepts. Moreover, we expect that diseases that inherently transmit with more difficulty, such as syphilis, will experience more frequent die-outs (especially in smaller cities) and prevalence levels will also be subject to higher levels of variation due to larger stochastic effects on the spreading dynamics of the disease. 

Furthermore,  Table \ref{table:NB_chl2} suggest that the socio-economic landscape in which STDs spread also affects the scaling patterns we infer. In fact, in environments that are more auspicious for STD transmission, the scaling exponents (intercepts) tend to be lower (higher).  

Consequently, we can hypothesize  that the lower (higher) the transmission potential of a disease, due to its inherent transmissibility capacity or due to a less auspicious environment,  then: 1) the more (less) its transmissibility appears to be contingent on population size, 2) the lower (higher) its overall incidence, and  3) the larger (lower) its variation in incidence. These inferences, although sensible and intuitive, were duduced from the statistical patterns of three STDs, and it is therefore  reasonable to be skeptical of their validity for other STDs, let along other infectious diseases or other contagion processes. To test these hypotheses we turn to, first, a mathematical models of infection, that informed by the empirical evidence in \cite{schlapfer2012scaling, tizzoni2015scaling},  can help us study the effect of  specific mechanistic pathways. Second, we provide an alternative yet related explanation to these observations through the prism of new theory of urban scaling.

\subsection{Mathematical epidemiological models}

\subsubsection{Deterministic model}

The SIS  (Susceptible-Infected-Susceptible) epidemic model is widely used to study the dynamics of  sexually transmitted diseases in homogeneously  \cite{KeelingRohani2008} and heterogeneously mixed populations, \cite{castillo1997effects, li2003coexistence, castillo1999competitive}. It is a simplified representation of an endemic disease with no lasting immunity. Therefore, this modeling approach, albeit simple, could be suitable for explaining some of the scaling patterns observed in the data of bacterial STDs. 

 In this framework, individuals are classified based on their infectious status as susceptible ($S$) or infected ($I$). Individuals become infected at rate $\beta$ and 
 recover at rate $\gamma$, where $1/\gamma$ is the expected time to recover. The transmission rate is defined as the per-capita contact rate, $c$, multiplied by  the probability a contact with an infectious individual leads to an infection, $p$. That is $\beta=c\times p$. Intuitively, it is clear that the number of contacts experienced by a random individual in a population depends on the size of the population, $N=S+I$, and and presumably, $c$ should increase with population size  \cite{sattenspiel2009geographic}. 

 Without making any assumptions about the functionality of $\beta(N)=c(N)\times p$, and considering a relatively short  modeling time frame (so that we can neglect demographic aspects such as births and deaths), the system describing the disease dynamics  is:  
\begin{eqnarray}
\frac{dS}{dt}&=&-\beta(N) SI/N +\gamma I \label{eqs2}\\
\frac{dI}{dt}&=&\beta(N) SI/N-\gamma I. \label{eqi2}
\end{eqnarray}
 
 This system has a basic reproductive number \cite{chowell2009basic} given by $R_0(N)= \beta(N)/\gamma$, and the endemic equilibrium can be found by solving $dI/dt=0$, yielding  
 \begin{equation}
 I^*(N)=N[1-1/R_0(N)]. \label{sim_endemic} 
\end{equation}

In the simplest case, we can assume that $c$ is constant, and therefore $R_0(N)=R_0$ (often suitable when $N \to \infty$). In this scenario, $I^*(N)$ is a linear function of $N$; hence that assumption does not lead to superlinear scaling.

  Building upon the empirical evidence that indicates that  social connectivity scales superlineraly with population size   \cite{schlapfer2012scaling, tizzoni2015scaling}, we can assume that the average number of contacts per individual, $\langle c \rangle$ also scales with population size as $\langle c \rangle \sim N^{\psi}, 0<\psi<1$, and the empirical estimations of $\psi$ fluctuate around 0.15 (0.11 to 0.2) depending on level of aggregation and country.

 Postulating that the contact rate, $c(N)$, is a function of population size given by 
\begin{equation}
c(N)=bN^a  \label{pl0}
\end{equation}
where $a$ and $b$ are positive constants. The expression of the transmission rate, $\beta(N)=c(N)\times p$, has one component which is affected by population size, i.e., $c(N)$, and $p$, which depends on the intrinsic transmissibility of the disease but also on sexual behaviors (e.g., use of condoms).\footnote{Noteworthy, the baseline parameter $b$ can be a function of socio-economic conditions. For example, cities with a large fraction of high-risk individuals (i.e., individuals with a large number of sexual partners) would feature a larger $b$. Also, since the mode of transmission of the disease (e.g., sexual, airborne) is crucial in defining what constitutes a contact, this can also affect the value of $b$. Additionally, we note that the socio-economic context  can affect not only  the infection risk $p$ (e.g., via better sex education and easy access to condoms), but it can also affect the recovery rate parameter $\gamma$. For instance, higher income is associated with better access to health care and treatment, both of which are critical in determining the speed of recovery. Thus, improved socio-economic conditions can reduce $1/\gamma$. }  

Substituting expression \eqref{pl0} in expression \eqref{sim_endemic} yields

  \begin{equation}
 I^*(N)=N\left[1- \frac{1}{\frac{p b}{\gamma} N^a}\right]. \label{istar}
\end{equation}

To understand the impact of the different parameters (i.e., $p, b, \gamma$) on the scaling of disease prevalence with population size\footnote{ If we assume that the system under study has reached its equilibrium, then the cumulative incidence from time $\tau_1$ to $\tau_2$ is equal to prevalence $I^*$ times a constant. More specifically, $Incidence(\tau_1, \tau_2)=\gamma (\tau_1-\tau_1)I^*$. Hence, at equilibrium the scaling properties of prevalence and incidence are equivalent.}, we first need to rewrite  expression \eqref{istar} in terms of $\log(I^*)$ and $\log(N)$.  Taking logs on both sides of   \eqref{istar} and after some transformations we obtain
  \begin{equation}
 \log(I^*)=\log(N) +\log \left(1- \frac{ e^{-a\log(N)}}{\frac{pb}{\gamma}} \right). \label{istarlog}
\end{equation}
 Taking partial derivatives in \eqref{istarlog} results in 

  \begin{equation}
 \frac{\partial \log(I^*)}{\partial \log(N)}=1+\frac{a}{\frac{pb}{\gamma} e^{a \log(N)}-1} =1+\frac{a}{R_0(N)-1} \label{deri*}
\end{equation}

In  cases where the disease can actually spread, that is, if $R_0>1$, the denominator in \eqref{deri*} is positive and so the derivative is larger than 1, which is indicative of the superscaling of incidence. To see this, note that based on equation \eqref{logyn}, the derivative $\partial \log(Y)/ \partial \log(N)=\alpha$, which in combination with  \eqref{deri*} renders the following relation:

  \begin{equation}
\alpha =1+\frac{a}{R_0(N)-1} \label{deri_alpha}.
\end{equation}
  
  From this expression we see that when $a=0$ (that is, when the contact rate does not increase with population size), the scaling exponent becomes 1 (linear).  
  
 Expression \eqref{deri_alpha} allows us to determine the effect of the different parameters in the scaling properties of disease prevalence. As $R_0(N)=\frac{p b}{\gamma} N^a$ increases the effect of population size on the scaling of  prevalence is reduced. In other words, if the disease becomes more transmissible-- either due to becoming more infectious (larger $p$), or due to becoming harder to detect and treat (larger $1/\gamma$), or due to sexual behaviors changing as to increase the rate of sexual contacts in the population (larger $b$)-- then the effect of population size in the contagion process is diminished. Conversely,  as the disease becomes less transmissible,  the effect of population size in contagion increments.

Moreover, the observation that larger scaling exponents correlate with lower intercepts (i.e., prevalence) can also be explained by differences in transmissibility of the STDs ($p$) combined with increased connectivity in larger populations (expression \eqref{pl0}). That is, as $R_0$ increases, the corresponding values of $\alpha$ decrease while the prevalence increases (see \eqref{sim_endemic}).  

\begin{figure}[ht]
\centering
 \includegraphics[width=4in, trim = 1in 2in 0in 1in,clip=true]{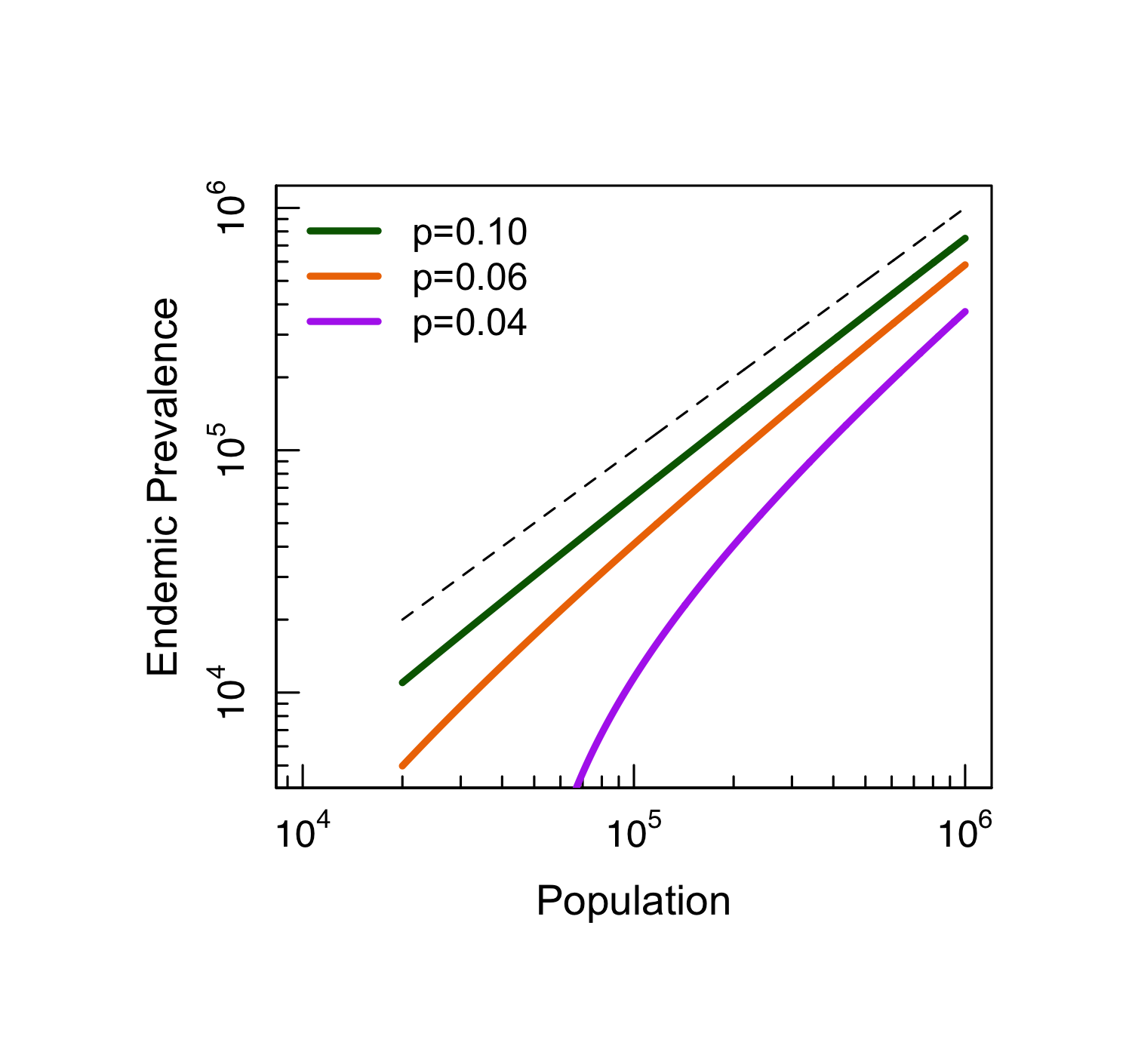}
\caption{\small Disease prevalence versus population size, for different intrinsic transmissibility values ($p$). As $p$ increases, the marginal effect of the population size decreases (i.e., lower scaling exponent). Other parameters: $a=0.15, b=0.5, \gamma=1/10.$ Note the similarity with left panel in Figure \ref{NB_chlam}.}
\label{SIS_pop_SE}
\end{figure}

To visualize these findings, Figure \ref{SIS_pop_SE} shows the plot of three curves corresponding to three diseases with different infectiousness $p$. From the figure we can readily observe that the  effect of population size on prevalence decreases as $p$ increases, while the prevalence levels increases with $p$.

\begin{figure}[ht]
\centering
 \includegraphics[width=0.7\textwidth, trim = 0in 0.3in 0in 0.5in,clip=true]{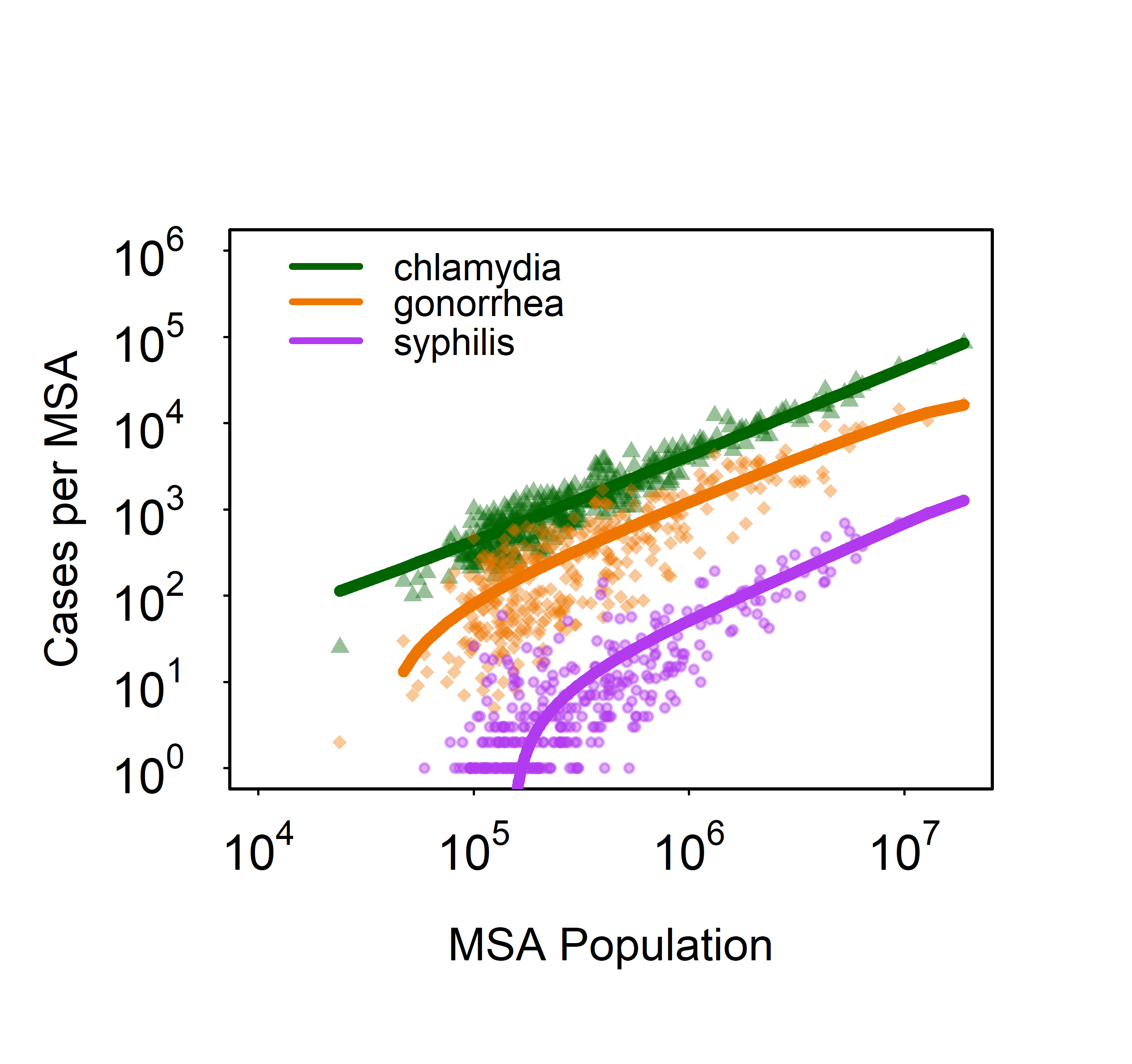}
\caption{\small Incidence of chlamydia, gonorrhea, HIV and syphilis versus MSA population, fitted with a loess line. Note the similarity with Figure \ref{SIS_pop_SE}.}
\label{SIS_pop_loess}
\end{figure}

 Finally, note that expression \eqref{deri_alpha} indicates that  the derivative $ \frac{\partial \log(I^*)}{\partial \log(N)}$ depends on $N$, unlike the case in \eqref{logyn} where $\alpha$, by design, is not a function of $N$. However, as Figure \ref{SIS_pop_loess} shows, fitting a smooth function (i.e., loess regression, a non-parametric method that is in principle agnostic to the shape of the data) to the log of the incidence versus log of the population (instead of a straight line) reveals that the relationship between of the log incidence and log population is only linear for relatively high populations, with a non-linear decaying  for lower populations. Remarkably,  using a simple SIS model this is precisely the behavior we obtain. Moreover, this behavior at low population sizes implies that changes in the environment can  have a relatively large impact in small populations.

\subsubsection{Stochastic model of contagion}

A key distinction between Figures \ref{SIS_pop_SE} and \ref{SIS_pop_loess} is the substantial variation seen in the empirical data versus the lack of it seen in the simulations from the deterministic model. The reason is that the process of disease transmission is inherently stochastic rather than deterministic. Hence, we now explore a stochastic version of the SIS model presented in the previous section. This will allow us to test whether diseases that transmit with less ease will also feature higher levels of variation at a given population size, as seen in the data. 

In this scenario, we again assume that the contact rate $c$ increases with population size as $c=bN^a$, the transmission rate is given by $\beta=p c$, where $p$ is the per-contact infection risk and $c$ is the number of contacts per unit time. The probability a susceptible individual becomes infected between time $t$ and $t+dt$ is given by 
\begin{equation}    
\mathbb{P}_I(t)=1-\exp[-\beta I(t)dt]=1-\exp[-pbN^a I(t)dt], \label{probInfhomo}
\end{equation}

where $I(t)$ is the number of infected individuals in the population at time $t$, and $N$, the population size.

\subsubsection{Simulations}

\begin{figure}[ht]
\centering
 \includegraphics[width=4in, trim = 1in 1in 0in 1in,clip=true]{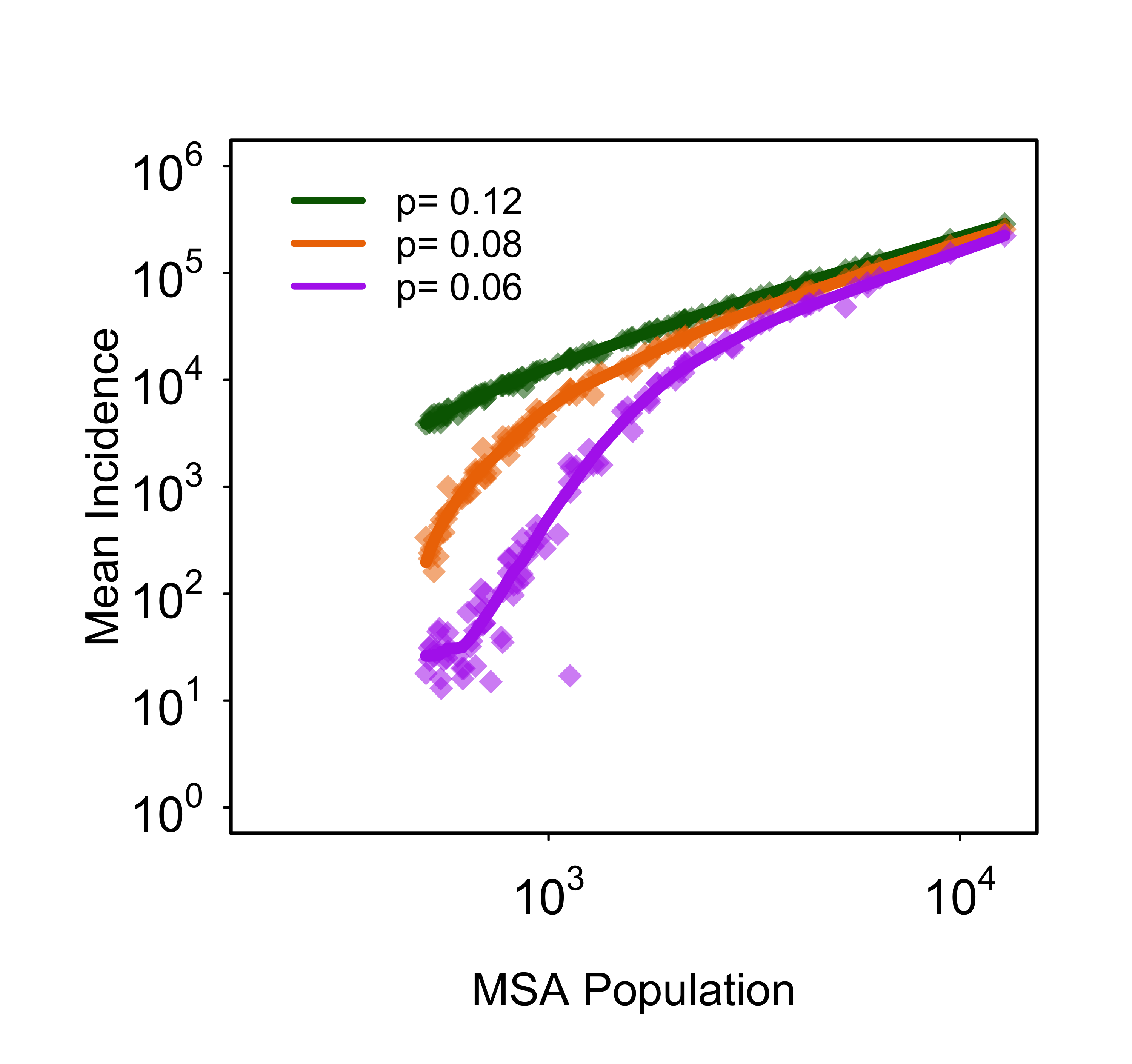}
\caption{\small Incidence versus population, fitted with a loess line. Mean of ten stochastic simulations for each population size, with $\gamma=0.3$ and $dt=0.1$.}
\label{SIS_stoch_pop_loess}
\end{figure}

Figure \ref{SIS_stoch_pop_loess} shows the results of the  SIS stochastic simulations. It shows a similar pattern to that of the empirical data in Figure \ref{SIS_pop_loess} and the deterministic SIS in Figure \ref{SIS_pop_SE}. Again, as the transmission rate increases, the scaling patterns become more linear, with a higher scaling exponent. Moreover, we also see in the figure that larger transmission rate are associated with lower variability, as seen in the empirical data. However, the variability in the simulations is considerably less.

This stochastic model helps characterize, for a given phenomenon, the range of variation due to pure randomness. In other words, when one wonders whether a city has an unusually high incidence by pure bad luck, one wants to have an expectation for how (un)lucky can a city get. This stochastic version of the model suggests that the wide variation in the data is not simply pure randomness, but rather that the variation itself has a structural origin, which can be understood from the lense of this SIS  model.

\subsection{New scaling theory}

\begin{figure}[h!]
\begin{center}
\includegraphics[width=4in, trim = 0in 0in 0in 0in]{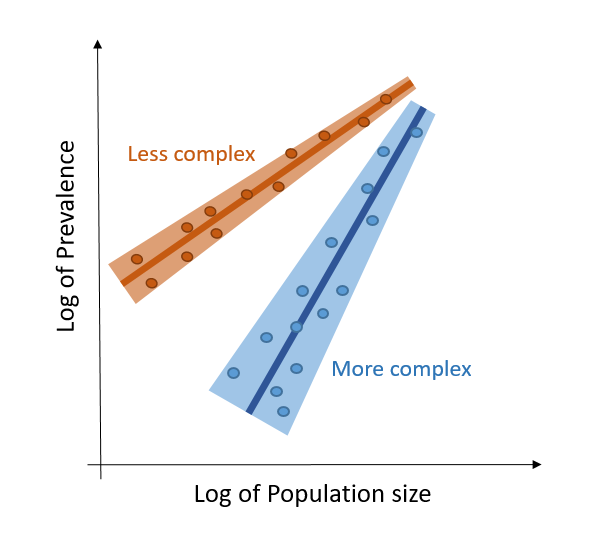}
\caption{\small Depiction of the scaling patterns of two different urban phenomena. Compared to the less complex phenomenon (in orange), the more complex phenomenon (blue) scales more steeply with population size (larger slope), features larger average prevalence (solid line) and larger variance across cities of similar size (wider bands).}
\label{Fig_1}
\end{center}
\end{figure}

We recently introduced a novel theory of urban scaling to explain the differences in, and relationships between, prevalence, scaling exponents, and cross-sectional variance of not only the incidence of STDs, but of a wide and diverse set of urban phenomena (e.g., education, employment, innovation and crime)\cite{gomez2016explaining}. This was achieved by elegantly coupling ideas from the fields of economic complexity (i.e., any given social phenomenon can only occur if a number of different, but complementary, factors are simultaneously present) \cite{HidalgoHausmann2009,HausmannHidalgo2011} and cultural evolution  (i.e., the diversity of factors grows logarithmically with population size) \cite{Henrich2004Tasmanian,henrich2015secret}. The main results can be summed up as: compared to a less complex phenomenon, a more complex phenomenon is expected to be less prevalent (rarer), scale more steeply with population size (more super-linear), and show larger variance across cities of similar size (noisier). See Fig \ref{Fig_1} for a depiction of these observations, and Figure 1 of reference \cite{gomez2016explaining} for more empirical examples.

A key conceptual contribution of this theory is giving center stage to the notion of ``complexity'' in order to shed light on our understanding of the scaling patterns of a myriad urban phenomena. Specifically, we posit that, to better understand these patterns, we should focus on the fact that social phenomena are the result of \emph{multiple} elements (e.g., material, social or biological) co-occurring in a social environment following a specific ``recipe''. A `complex phenomena' in this setting is one that, in order to occur, requires the presence of a large number of complementary factors; or put differently, compared to a less comples one, a more complex phenomenon has a lower probability of occurrence. Importantly, as we will see below, the complexity of a phenomenon is determined by 1) its \emph{inherent} difficulty (e.g., disease transmissibility in the context of STDs), 2) the social fabric (e.g., network of interactions between individuals) in which the phenomenon occurs/spreads and 3) the capacity of the individuals involved to contribute to the ocurrence of the phenomenon (e.g., awareness and exercising of methods to prevent the spread of infections). 

\subsubsection{The three facets of complexity}

A central mathematical relation of the theory derived in \cite{gomez2016explaining} states that:

\begin{equation}
E[Y_{c,f}]=Ne^{-M_f q(1-r_c)} \label{e1}
\end{equation}

where $E[Y_{c,f}]$ is the expected number of individuals engaged to a given phenomena $f$ (e.g., becoming infected with syphilis) in a city of population size $N_c$. The parameter $M_f$ is the number of necessary elements a given urban activity (e.g., transmission of an infectious disease) needs in order to occur. The parameter $q$ is the probability any given individual needs any of the $M_f$ elements to be counted in said activity and it is associated with the  lack of susceptibility of individuals (e.g, to become infected). Lastly, $r_c$ represents the diversity of elements city $c$, with size $N_c$, provides and, more precisely, it is the probability that an individual encounters any given element while living in the city. Thus, $r_c$ quantifies the size of the social, economic, and cultural repertoire that is available in a city and to which individuals are exposed to.

The link between Equation \eqref{e1} and the power-law expression in Equation \eqref{yn} emerges when $r_c$ grows logarithmically with population size as $r_c(N_c)=a+b\ln(N_c)$. This logarithmic scaling of urban diversity, we posit, is a result that emerges from a selection and cumulative process of cultural evolution \cite{Henrich2004Tasmanian,henrich2015secret}. Substituting this functional form in Equation \eqref{e1} yields:

\begin{equation}
E[Y_{c,f}]=e^{-M_fq(1-a)} N_c^{M_fqb+1}. \label{epercapita}
\end{equation}

Comparing expression \eqref{epercapita} with the standard scaling model, $Y=Y_0 N^\beta$, the baseline prevalence $Y_0$ is given by $e^{-M_f q(1-a)}$, the scaling exponent $\beta$ by $M_fqb+1$ (see \cite{gomez2016explaining} for more details). Note also the interesting similarities between expression \eqref{epercapita} and the statistical model in  \eqref{interact}. 

The exponent in \eqref{e1} has three terms that are proportionately related to the \emph{complexity} of the phenomenon or activity, in the sense that increasing any of the terms decreases the probability that any given individual is engaged in said phenomenon:
\begin{itemize}
\item 1) the \emph{inherent} complexity of the phenomenon, which is represented by $M_f$
\item 2) the lack of susceptibility of individuals to be engaged with the phenomenon, represented by $q$
\item 3) the unsuitability of the system/environment for the phenomenon to occur, given by $1-r_c$. 
\end{itemize}

\subsubsection{Case example: scaling of two sexually transmitted diseases}

\begin{figure}[h!]
  \centering
  \begin{minipage}[b]{0.49\textwidth}
    \includegraphics[width=\textwidth]{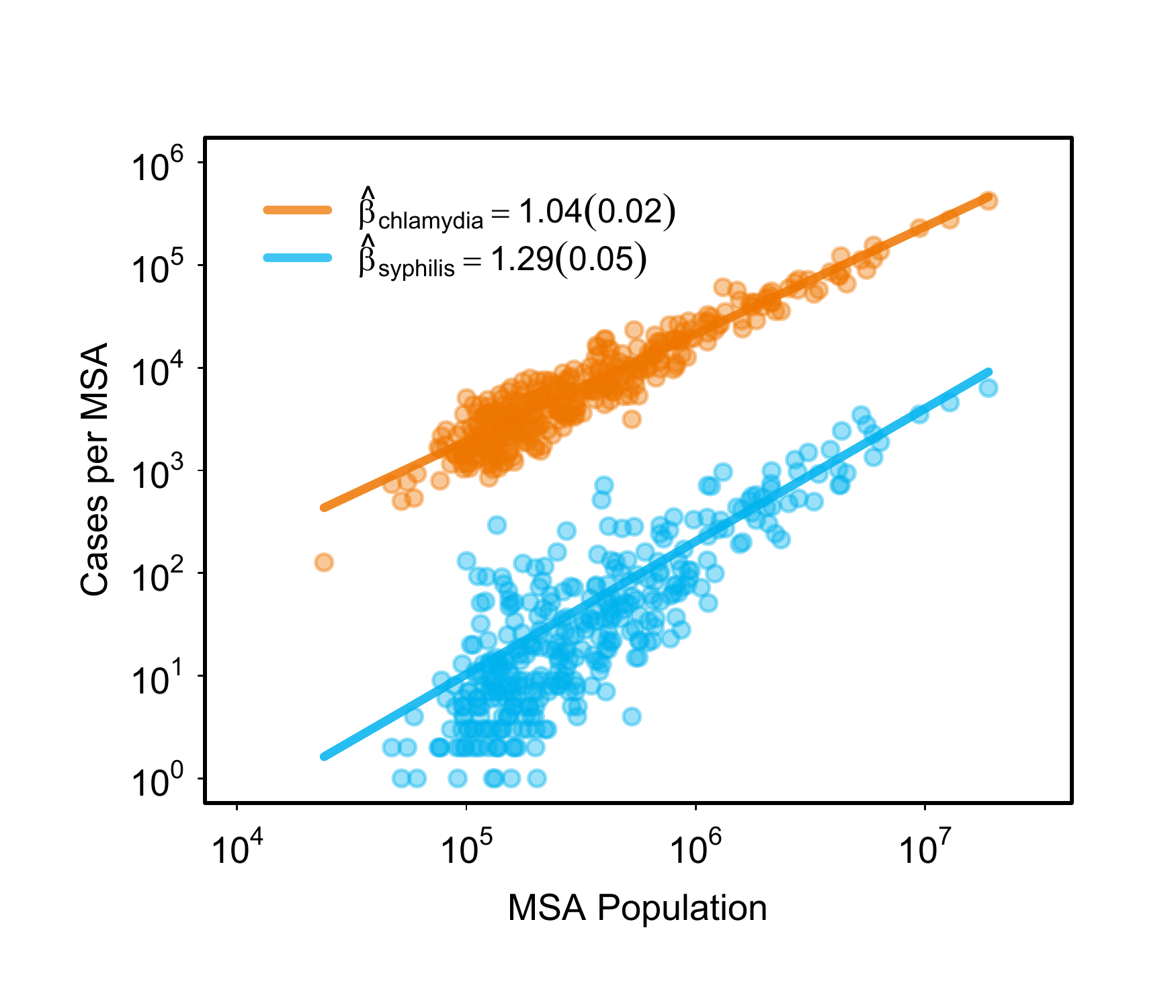}
  \end{minipage}
  \begin{minipage}[b]{0.49\textwidth}
    \includegraphics[width=\textwidth]{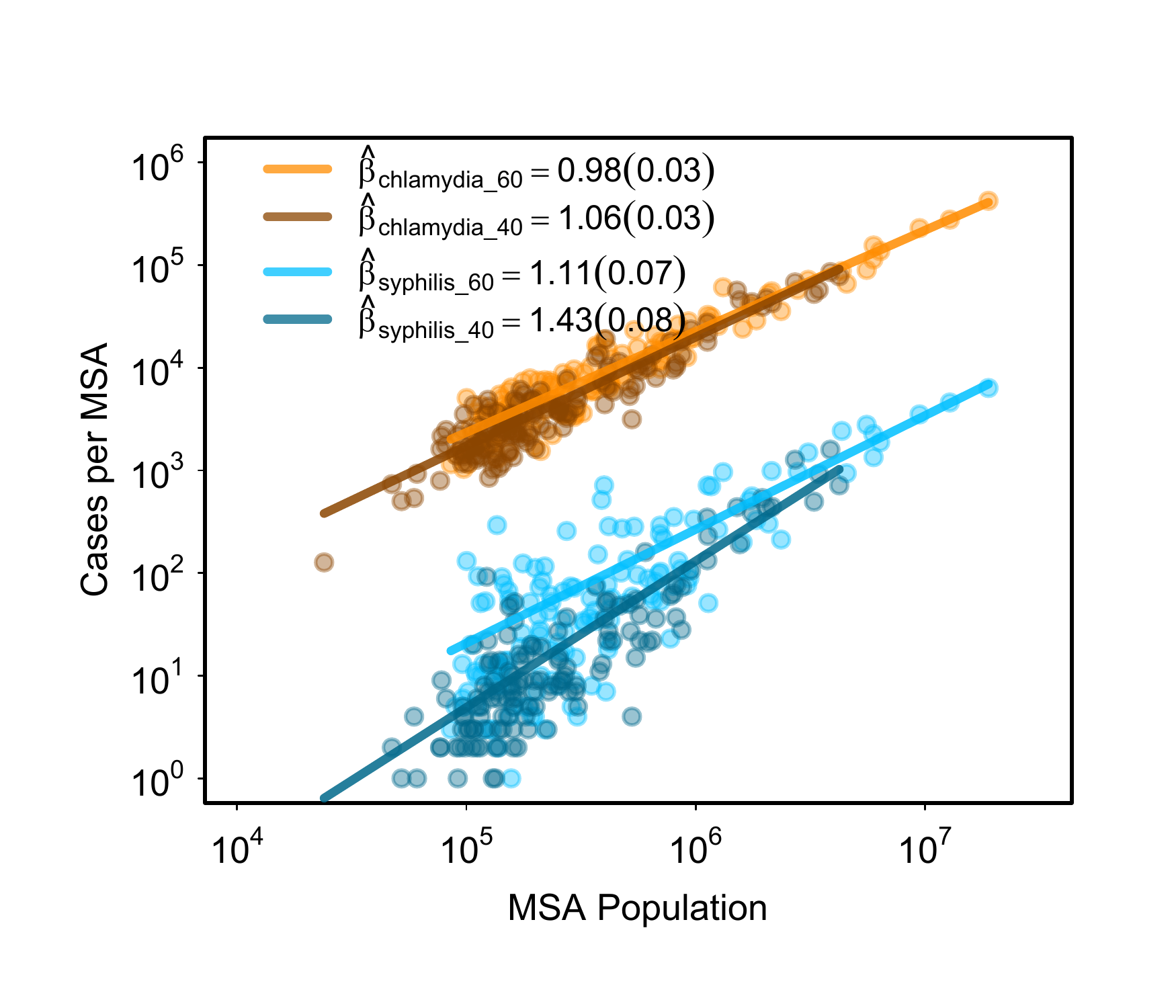}
  \end{minipage}
    \caption{\small (left) Scaling patterns of the 5-year average incidence of STDs stratified by disease (chlamydia [orange] and syphilis [blue]) in metropolitan statistical areas (MSA) in the US. (right) Scaling patterns of STDs stratified by disease and by gini index of cities (cities with high income inequality [light colors] and cities with low income inequality [dark colors]). These two groups cities by inequality are defined, respectively, as those that belong to the two highest quintiles or the top 40th percentile (light blue) and the two lowest quintiles or the bottom 40th percentile (dark blue) in terms of the gini coefficient. The legends show the corresponding scaling exponents and their standard errors.}
  \label{Fig_2}
\end{figure}

Here we illustrate these concepts and the link of this theory to the empirical evidence discussed above regarding syphilis and chlamydia. Fig \ref{Fig_2}, left panel, displays the three hallmarks described above: compared to syphilis, chlamydia has higher prevalence, lower scaling exponents ($\beta$) and lower variability. We hypothesize that a key driver of the diverging scaling patterns of these two STDs is the difference in their inherent transmissibility. Chlamydia is more infectious and less symptomatic than syphilis, hence the spreading capacity of chlamydia is considerably higher than that of syphilis. This observation is consistent with our theory once we make the connection between complexity and (lack of) transmissibility. Syphilis is more complex than chlamydia in the sense that syphilis requires more elements ($M$ in the model) to be transmitted than chlamydia does. For example, given its lower infectiousness per sexual contact compared to chlamydia, syphilis requires, on average, more sexual interactions for a transmission event to occur; and since it is more symptomatic, it requires for people to be more negligent about their sexual behavior and health.

The right panel of Fig \ref{Fig_2} shows how the scaling properties of syphilis and chlamydia are not only disease-dependent, but also context-dependent (note that  Fig \ref{Fig_2} represents a visual confirmation of the observations derived from the results in Table \ref{table:NB_chl2}). In this example, the ``context'' is determined by the level of income inequality measured by the gini coefficient. Focusing on the case of syphilis, the prevalence of syphilis is higher while its scaling exponent and variability are lower in cities with higher levels of income inequality (i.e., cities belonging to the two highest quintiles with respect to their gini coefficients) than in cities with lower levels of income inequality (i.e., cities belonging to the two lowest quintiles). Similar divergent patterns are observed if cities are instead categorized by the other covariates presented in the previous section. All of these metrics are a (partial) reflection of socio-economic aspects that affect the properties of the sexual networks on which the disease is spreading, and the right panel of Fig \ref{Fig_2} shows that the incidence of syphilis is higher, with a lower scaling exponent and variability, in a more enabling system. In the terminology of our theory, the scaling of syphilis depends on the suitability of the system ($r_c$ in the model) to spread it.

In sum, the left panel of Fig \ref{Fig_2} shows how, everything else constant (e.g., system propensity), a difference in the inherent complexity of two phenomena (e.g., disease transmissibility) can affect, in very specific ways, how these phenomena scale and vary with population size. The right panel demonstrates that, controlling for the inherent complexity of the phenomenon, the propensity of the environment (e.g., level of income inequality)  determine similarly the scaling patterns of said phenomenon. Moreover, for relatively less complex phenomena, like becoming infected with chlamydia, the effect of the system suitability (in this case, the propensity of the system to facilitate the spread of chlamydia) on the scaling pattern is more limited, as it is demonstrated by the more similar patterns of chlamydia in cities with large and small levels of income inequality.

Unfortunately this data does not contain information at the individual level, therefore we cannot use it to test one of the predictions of our theory, namely, that the susceptibility of individuals has an effect on the scaling patterns similar to that of the inherent transmissibility and system's propensity. 

This model is in line with the previous framework (epidemiological model). It states that the three aspects driving the incidence of a disease (inherent complexity, system propensity and individual susceptibility) are not independent from one another, but rather intertwined in potentially convoluted ways. For example, the propensity of a city to the spread of an STD is tightly linked to how susceptible its inhabitants are, which also affects the person-to-person transmissibility of the disease. However, this example also demonstrates that decomposing complexity into these three components is very useful in advancing our understanding the drivers of scaling patterns.

\subsubsection{Complexity and predictability of phenomena}

A recent work by Scarpino et al. \cite{scarpino2017predictability} on the predictability of infectious diseases, indicates that the time seriers of diseases with lower $R_0$ (i.e., lower infection potential) are, on average, harder to predict. Moreover, the authors showed that diseases that spread in networks with topologies that are less conducive to disease spread are also less predictable. These findings make sense intuitively given that \emph{rarer} events are tipically more difficult to predict. Interestingly, these findings are connected to the insights put forth in this paper.

To elucidate the connection we need to note three other links. First, predictability, as defined in \cite{scarpino2017predictability} (using permutation entropy as a model independent measure of predictability), is related to Kolmogorov complexity \cite{politi2017quantifying}, which is the length of the shortest computer program that produces the object (phenomenon) in question. Second, in our framework, more complex phenomena are, by definition, those with more necessary elements for the phenomena to occur. Lastly, more complex phenomena feature more \emph{cross-sectional} variability and, consequently, the occurrence of more complex phenomena are more difficult to predict, which closes the explanatory loop with the findings in \cite{scarpino2017predictability}. It is also interesting to note how the cross-sectional variability of disease spread is lower both when diseases have lower infection capacity (intrinsic property of the phenomenon) and when diseases spread in less conducive environments (e.g., social networks), which is nicely tied in to how we dicompossed the concept of complexity herein.      

The intrinsic complexity of a disease not only affects the  \emph{cross-sectional} (i.e., in cities of a given population size) predictability of the disease spreading process, but also its predictability in \emph{time} (i.e., a given diasese in a city). To show this, we use permutation entropy, akin to \cite{scarpino2017predictability}, to measure the predictability (defined as 1-permutation entropy) of infection time series for the three STDs in the dataset.  

As a complementary way to measure the variability of the time series of the three diseases across cities, we also computed the coefficient of variation (COV=standard deviation/mean) of each disease in each city.

\begin{figure}[h!]
  \centering
  \begin{minipage}[b]{0.49\textwidth}
    \includegraphics[width=\textwidth]{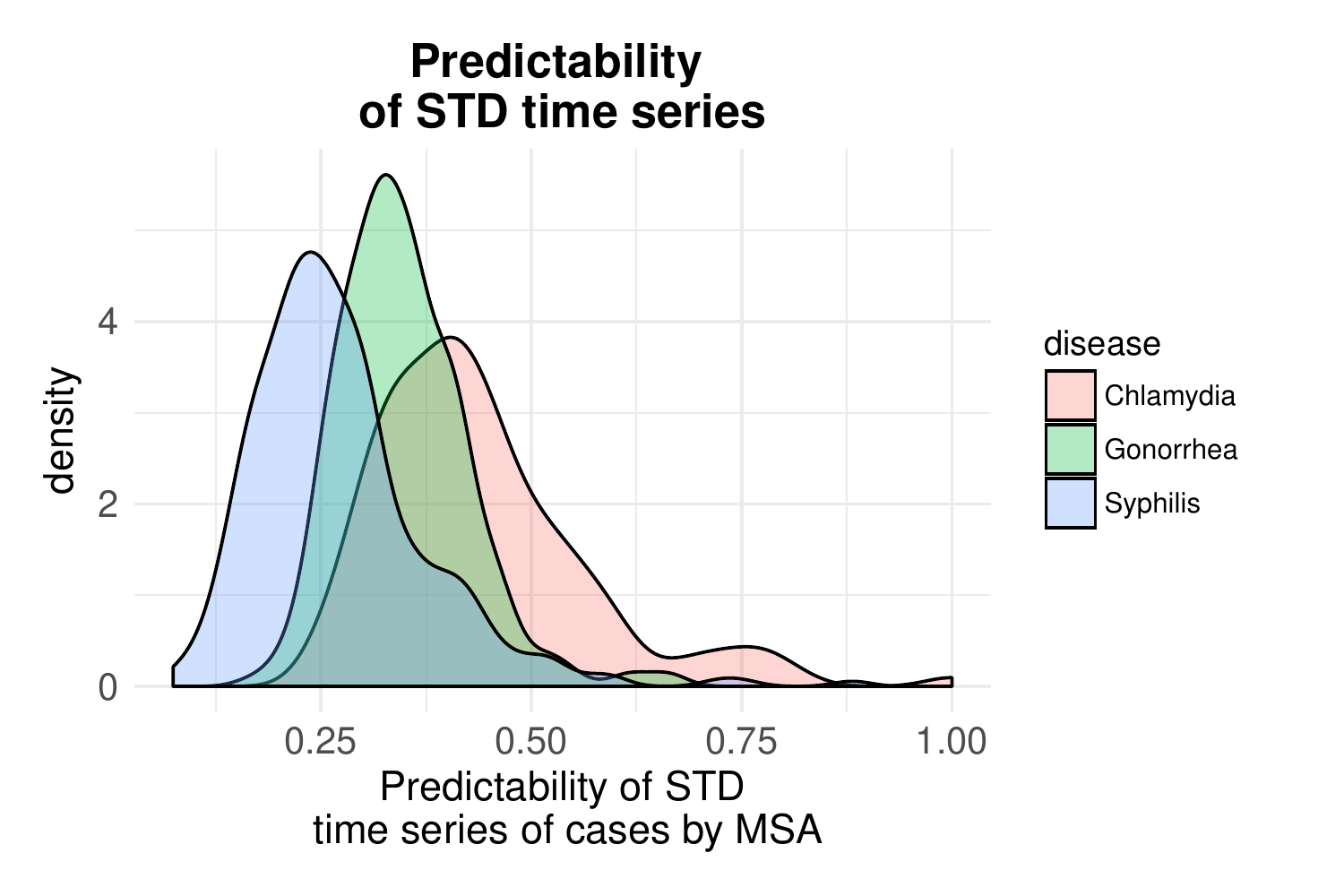}
  \end{minipage}
  \begin{minipage}[b]{0.49\textwidth}
    \includegraphics[width=\textwidth]{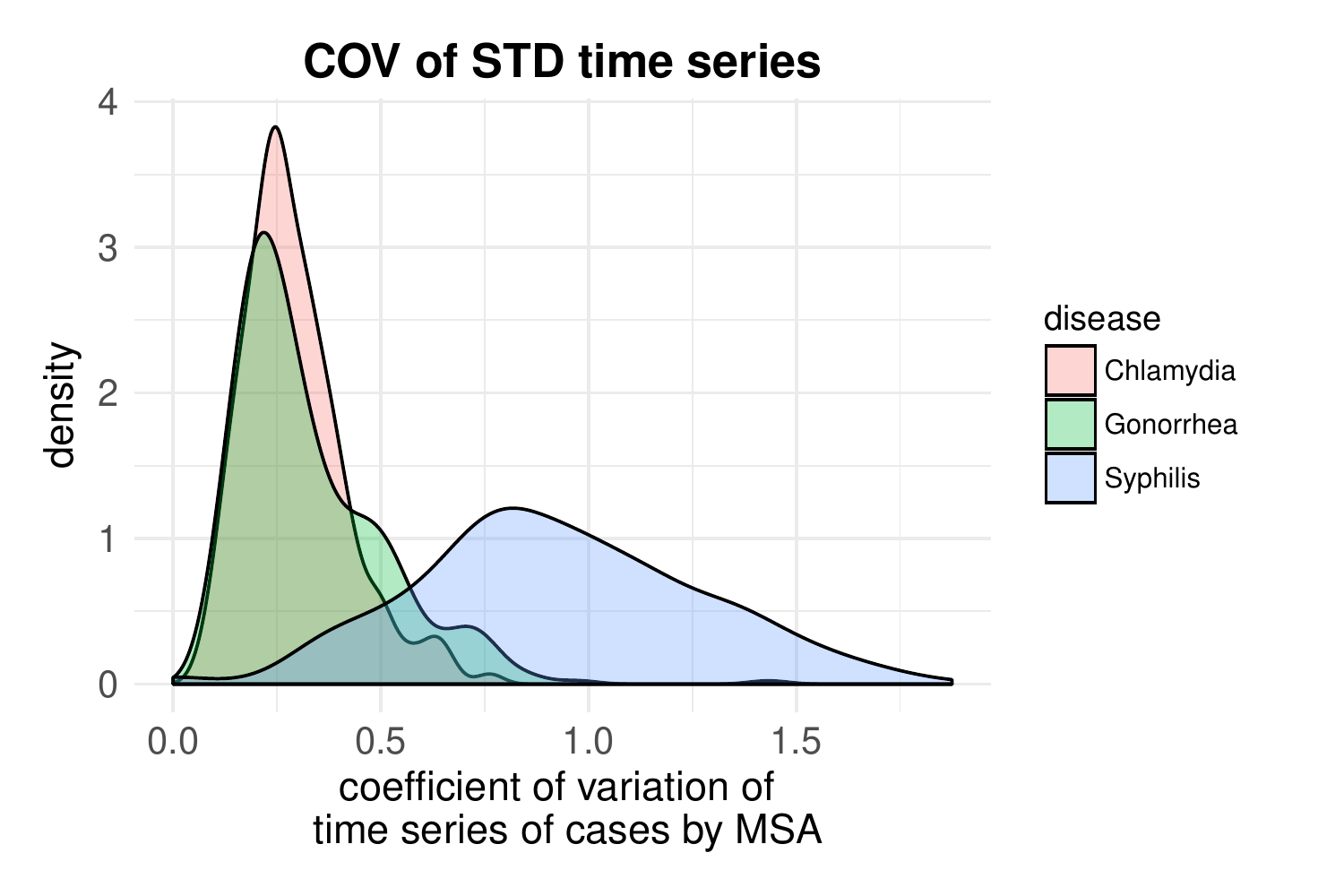}
  \end{minipage}
    \caption{\small Distributions of predictabilities (1-permutation entropies) (left panel) and COV (right panel) for each diasese across cities. We compute the permutation entropy of the time series of each disease in each city with at least 10 years of annual incidence data (to ensure a more robust/data-rich measure of permutation entropy). The chlamydia, gonorrhea and syphilis time series data span from 1996, 1995 and 1984, respectively, until 2011. We used the `statcomp' package in R, and a sliding window size of 4 (we also tried window sizes of 3 and 5 and the results were similar).}
  \label{Fig_densities_perent}
\end{figure}

Figure \ref{Fig_densities_perent} shows the distributions of predictabilities (left panel) and COV (right panel) for each diasese across cities. These figures show that for diseases with lower infection capacity (e.g., syphilis), the time series predictability of the spreading process is lower and the variability is larger, in line with the findings in \cite{scarpino2017predictability} and with what we noted before regarding cross-sectional variability.

\section{Discussion}

The empirical evidence present us with an image in which the statistical patterns of STD incidence in cities are systematically shaped by the intrinsic and contextual factors that determine the spreading capacity of the pathogen. Specifically, the three more salient features of the scaling patterns, namely the scaling exponent, the baseline prevalence and the cross-sectional variance of the incidence, vary in a predictably way with the inherent infectiousness of the disease and the socioeconomic properties of the cities in which they spread. Diseases that spread more easily are associated with lower scaling exponent, higher prevalence and lower variance. 
 
Going beyond the statistical ascertainment and description of these interesting empirical relationships, we offered two mechanisitc models that provided compelling explanations for these patterns and their underlying drivers. In the first approach, we used a simple model of infectious disease spread (SIS model) that incorporated in its mathematical formulation the empirical observation that contact rates increase with population size. This model, in its determinitic and stochastic versions, helped us confirmed that, indeed, the scaling patterns of diseases with higher transmission rates have lower scaling exponent, higher prevalence and lower variance. Additionally, this model also captured the non-linear (in the log-log scale) decay of the incidence with population size for cities in the lower end of the population size spectrum. This behavior is not easily obtained with traditinal scaling models, but it emerged remarkably naturally from this simple model. 

In the second approach, we used a new scaling theory that was built from first principles and effectively coupled ideas from economic complexity and cultural evolution. Through the prism of this theory we were  able to provide coherent and intuitive explanations of these empirical observations. Specifically, the structure of the theory's main  mathematical relationship revealed a very insighftul conceptual dissection of the scales at which the different facets of the``complexity" of a phenomenon operate. The theory suggested that the \emph{net complexity} of a phenomenon results from the multiplicative effects of the inherent complexity of phenomenon, the level of engagement of the individuals involved, and the conduciveness of the surrounding environment for the phenomenon to occur. Translated to the context of infectious diseases, these three aspects can be intuitively interpreted as the intrinsic infectiousness of the disease, the susceptibility of individuals and the suitability for disease spread of the underlying social fabric, respectively.  These factors act at different levels (i.e., intra individual, inter individual and community levels), but are interrelated. The first factor is tied in to the biology of the pathogen and the host, with the risk of infection per sexual act varying also with the mode of transmission. The second term relates to individuals' risk behaviour (e.g., condom use, number of sexual partners) and perceived risk, which could be in turn affected by educational level, and access to medical and counselling services, as well as the symptomatology of the disease. Lastly, the third term refers to aspects such as the sexual network in which individuals are embedded and the prevalence of the disease in the community, which is in turn governed by social and cultural norms and the socio-economic landscape of the community.

Inspired by this conceptual framework and the recent work in \cite{scarpino2017predictability}, we showed that the intrinsic complexity of a disease not only determines the cross-sectional variability of disease incidence, but also its temporal predictability.

 The connectivity of sexual networks \cite{liljeros2001web}, in combination with their size, can influence the transmissibility of infectious diseases \cite{watts1998collective,may2001infection,pastor2001epidemic,newman2002spread, keeling2005implications}. Arguably, the structure and size of high-risk sexual networks are affected by the size of the cities in which they exist. For example, larger cities can foster either larger, or more numerous, high-risk sexual networks. In other words, the higher  rates of STD prevalence in large cities could be explained, at least in part, by a disproportionate presence of high-risk groups in those populations. Therefore, one interesting avenue to improve the contagion model used here is to relax the homogenous mixing assumption and instead assume that individuals are embeded in a network of (sexual) interactions. In this setting, it would be interesting to explore how the network structure (e.g., degree distribution) affects the scaling patterns. In fact, based on the findings of these paper, we hypothesize that populations featuring network topologies that are less suitable for wide spread of infections (e.g., networks with low contact heterogeneity) will lead to scaling patterns of incidence characterized by high scaling exponents, low baseline prevalence and high crioss-sectional variability.  Additionally, the variability in the spreading process due to the network structure could help explain the amount of variance seen in the real world data that was not captured in the stochastic model. Within a network framework it would also be possible to investigate the effects of different modes of transmission (e.g., sexually transmitted or direct contant versus airborne) in the scaling patterns.

It is important to note that the structure of the networks will depend on the inherent transmissibility of the disease. For example, the network of contacts underlying the transmission of chlamydia (a largely asymptomatic disease) is likely to be larger and more homogenous than that of syphilis (a more symptomatic disease) in which spatial heterogeneities are larger \cite{ patterson2015per, chesson2010distribution}. Connecting this observation to the concept of complexity discussed above, it is helpful to partition the net complexity into the three intuitive facets described before; however, it is important to recognize that these three facets of complexity are inexorably interrelated.

The two modeling approaches proposed herein to explain the scaling patterns of STDs in cities are based in different premises and assumptions. The contagion model is a dynamic non-linear formulation of a spreading process in time, whereas the scaling theory is based on probabilistic formulations based on first principles and does not explicitely incorporate time. However, there are a number of interesting parallels between the two approaches. For example, $R_0$ and \emph{net complexity} are inversely related. We showed how both these metrics provide a convenient mathematical structure that lend itself to helpful conceptual descriptions of the drivers of these contagion processes.    

These insights can be translated into policy recommendations. The scaling analysis indicated that STDs that transmit with less ease feature a larger superscaling effect. Therefore, resources and efforts to contain the spread of less prevalent diseases should focus on larger cities. Interestingly, a disease containment effort that distributes resources uniformly accross all populations, instead of focusing on larger cities, may result in lower overall disease prevalence, but simultaneously lead to a scenario in which larger cities are even more disproportionately affected by the disease than before the containment effort came into effect. Another critical insight from these investigations was that a favourable socio-economic landscape has an effect in  the scaling patterns akin to the one that the intrinsic infectivity of a disease has. Therfore, if the scaling patterns of a given infectious disease in two sets of cities (e.g., cities within two geographic regions of the USA, or cities within two countries) differ substantially, we can at least hypothesize that the cities in which the scaling pattern had a lower exponent and higher prevalence have, on average, a more conducive environment for the spread of the disease. This type of information may be useful to a policy maker that is, for instance, trying to determine how to effectively deploy limited resources into two or more countries to contain the spread of a particular infectious disease.

\subsection*{Acknowledgements}

We would like to thank Julia Wei Wu for helpful discussions.

\small
\bibliographystyle{plain}
\bibliography{ID_scaling_v4_arxiv}

\clearpage
\newpage


\end{document}